\documentclass[sigconf,edbt]{acmart-edbt2019}

\usepackage{booktabs} 
\usepackage[utf8x]{inputenc}
\usepackage{graphicx}
\usepackage{algorithm}
\usepackage{algpseudocode}

\newcommand{\hide}[1]{}

\newtheorem{problem}{Problem}

\newlength\myindent
\usepackage{array}
\newcolumntype{L}[1]{>{\raggedright\let\newline\\\arraybackslash\hspace{0pt}}m{#1}}
\newcolumntype{C}[1]{>{\centering\let\newline\\\arraybackslash\hspace{0pt}}m{#1}}
\newcolumntype{R}[1]{>{\raggedleft\let\newline\\\arraybackslash\hspace{0pt}}m{#1}}

\begin{document}
\title{
Optimal Algorithm for Profiling Dynamic Arrays\\ with Finite Values
}

\author{
Dingcheng Yang, Wenjian Yu, Junhui Deng}
\affiliation{%
  \institution{BNRist, Dept. Computer Science \& Tech., }
  \city{Tsinghua Univ., Beijing, China} 
}
\email{ydc15@mails.tsinghua.edu.cn, yu-wj@tsinghua.edu.cn, deng@tsinghua.edu.cn}

\author{Shenghua Liu}
\affiliation{%
  \institution{CAS Key Lab. Network Data Science \&
Tech.,}
  \city{Inst. Computing Technology} 
  \state{} 
  \country{Chinese Academy of Sciences, Beijing, China}
  \postcode{43017-6221}
}
\email{liushenghua@ict.ac.cn}

\renewcommand{\shortauthors}{}

\begin{abstract}
How can one quickly answer the most and top popular objects at any time, 
given a large log stream in a system of billions of users? 
It is equivalent to find the mode and top-frequent elements in 
a dynamic array corresponding to the log stream.
However, most existing work either restrain the dynamic array within a 
sliding window, or do not take advantages of only one
element can be added or removed in a log stream.
Therefore, we propose a profiling algorithm, named S-Profile,
which is of $O(1)$ time complexity for every updating of the dynamic array,
and optimal in terms of computational complexity.
With the profiling results, answering the queries on the statistics of
dynamic array becomes trivial and fast.
With the experiments of various settings of dynamic arrays,
our accurate S-Profile algorithm outperforms 
the well-known methods, showing at least 2X speedup to the heap based approach and 13X or larger speedup to the balanced tree based approach. 


\end{abstract}

%
%


\maketitle
\section{Introduction}

Many online systems, especially with billions of users, are generating
a large stream of logs \cite{dietz2018big}, recording users' dynamics in the systems, e.g.~
users (un)follow other users, ``(dis)like'' objects, enter (exit)
live video channels, and click objects.
Then, a question is raised:

\emph{How can we efficiently know the most popular objects (include users), 
i.e.~mode, top-K popular ones, and even the distribution of frequency in
a fast and large log stream at any time?}

Mathematically, the questions converge to calculate and update the statistics
of a dynamic array of finite values. 
Thus the existing fast algorithms on the statistics are as follows:

\emph{Mode of an array.}
The mode of an array and its corresponding frequency can be calculated by sorting the 
array (if it's of numeric value) and 
scanning the sorted array in
$O(n \log n)$ time, where $n$ is the length of array
\cite{dobkin1980determining}. Notice that through judging the frequency of mode
we can solve the element distinctness problem, which was proven to have to be
solved with $\Omega(n \log n)$ time complexity
\cite{steele1982lower,lubiw1991lower}. Therefore, calculating the mode of an
array has the lower bound of $\Omega(n \log n)$ as well. If the elements of
array can only take finite values, the complexity of calculating the mode can
be reduced. Suppose they can only take $m$ values. One can use $m$ buckets to
store the frequency of each distinct element. Then, the mode can be calculated
in $O(n+m)$ time by scanning the $m$ buckets. 

The problem of range mode query, which calculates the mode of a sub-array $A[i
\dots j]$ for a given array $A$ and a pair of indices $(i,j)$, has also been
investigated \cite{krizanc2005range,petersen2009range,chan2014linear}. The
array with finite values was considered. With a static data structure, the
range mode query can be answered in $O(\sqrt{n/\log n})$ time
\cite{chan2014linear}. 
 


\emph{Majority and frequency approximation.}
The majority is the element whose frequency is more
than half of $n$. 
An algorithm was proposed to find majority in $O(n)$ time and $O(1)$ space
\cite{boyer1991mjrty}. Many work on the statistics like frequency count and
quantiles, are under a setting of sliding window
\cite{arasu2004approximate,datar2002maintaining,babcock2002sampling,gibbons2002distributed,lin2004continuously}. 
They consider the most recently observed data elements (within the window) and
calculate the statistics. Space-efficient algorithms were proposed to maintain
the statistics over the sliding window on a stream.

However, those existing work slow down their algorithms
without considering that the increase and decrease
of object frequency are always 1 at a time in log streams.
Therefore, we propose an algorithm S-Profile to
keep profiling the dynamic array. With such a profile, we can
answer the queries of the statistics: mode, top-K and frequency distributions.

In summary, S-Profile has the following advantages:
\begin{itemize}
    \item[-] \textbf{Optimal efficiency:} 
	S-Profile needs $O(1)$ time complexity and $O(m)$ space complexity to
	profile dynamic arrays, where $m$ is the maximum number of objects. 
    \item[-] \textbf{Querying Statistics:}
	With the  profiling, we have sorted frequency-object
	pairs, and can simply answer the queries on mode, top-K, majority and other statistics 
	in $O(1)$. 
    \item[-] \textbf{Applicable:} Our S-Profile can be plugged into most of log
	streams in many systems, and profiling objects of interest. 
\end{itemize}

%
In experiments, S-Profile 
compares with the existing methods 
in various settings of 
dynamic arrays, and shows its performance and robustness.

\section{An $O(1)$-Complexity Algorithm for Updating the Mode and Statistics}

We define tuples $(x_i, c_i)$ as a log stream,
where $x_i$ and $c_i$ is the object id and action in the $i$-th tuple. 
Action $c_i$ can be either ``add'' or ``remove'', which, for example, can indicate
object $x_i$ is ``liked'' or ``disliked'', or user $x_i$ is followed or
unfollowed.
Conceptually, we could imagine a dynamic array $A$ of objects associated with 
a log stream, by appending object $x_i$ into $A$ if $c_i$ is ``add'', and
deleting object $x_i$ from $A$ if $c_i$ is ``remove''.
Dynamic array $A$ is not necessarily generated and stored, which is defined for convenient 
description of our algorithm.

Therefore, our problem can be described as follows:

\begin{problem}[Profiling dynamic array]
\label{probstreamlog}
    \textbf{Given:} a log stream of tuples $(x_i, c_i)$ adding and removing an object each time,
\begin{itemize}
\item[-] \textbf{To find:} fast profiling of dynamic array $A$ of objects at any time,
\item[-] \textbf{Such that:} answering the queries on mode, top-K and other
    statistics of objects is trivial and fast.
\end{itemize}
\end{problem}




Let $m$ be the maximum number of distinct objects in a log stream 
or dynamic array $A$.
Without loss of generality, we assume id $x_i \in [1, m]$, i.e.~integers
between 1 and $m$. 
For any $m$ distinct objects, we can map them into the 
integers from 1 to $m$ as ids.

We can use $m$ buckets to store the frequency
of each distinct object. Let $F$ be such a frequency array
with length $m$. $F[i] \in F$ is the frequency of object with id $i$. 
With $F$, most statistics of $A$ can be calculated without
visiting $A$ itself. For example, the mode of $A$ refers to locations in $F$
where the element has the maximum value. 
    Although updating $F$ with each tuple of a log stream trivial and costs $O(1)$, 
    finding the maximum value in $F$ at each time is still time consuming.

    Therefore, we first introduce a proposed data structure of a profile, named  ``block
    set'', which can answer the 
    statistical queries in a trivial cost. And we show later that we can maintain
    such a profile in $O(1)$ time complexity and $O(m)$ space complexity.

\subsection{Proposed data structure for profiling}


In order to find the mode of $A$, we just need to care about the maximum in $F$. If only integers are added to $A$, the maximum element of $F$ can be easily updated. However, in Problem 1 removing integer is also allowed. This complicates the calculation of mode and other statistics of $A$. 
So, the sorted array $T$ must be employed and maintained. To facilitate the queries, $T$ can be implemented as a binary tree. The heap and balanced tree are two kinds of binary tree, and are widely used for efficiently  maintaining an sorted  array. 
Upon a modification on $A$, they both can be updated in $O(\log m)$ time. The root node of a heap is the array element with the extreme value. This means the heap is only suitable for producing the $A$'s elements with either maximum frequency (the mode) or the minimum frequency. The balanced tree is good at answering the query of median of $A$, and can also output the mode and top-K elements, etc.  
It should be pointed out, these general algorithms do not take the particularity of Problem 1 into account (the modification on $F$ is restricted to plus 1 or minus 1). By the way,  no one can maintain the sorted array under an arbitrary modification with a time complexity below $O(\log m)$, because it can be regarded as a sorting problem which has been proven to have $\Omega(m \log m)$ time complexity.

We use Figure 1 as an example to illustrate the proposed data structure for maintaining $T$. Suppose $T$ has frequency values in ascending order.
In order to locate the index of the $i$-th element of $T$ in $F$ and vice verse,
two conversion arrays are defined: $TtoF$ and $FtoT$. In other words, we have
$T[i] = F[TtoF_i]$ and $F[i] = T[FtoT_i]$. Here, we use both the subscript and
bracket notation to specify an element of array. 
As shown in Figure 1(c), we can partition $T$ into nonoverlapped segments according to its elements. Each such segment is called  \emph{block} here and represented by an integer triple  $(l, r, f)$, where  $l$ and $r$ are starting and ending indices respectively, and $f$ is the element value (frequency). So, a block $b=(l, r, f)$ always satisfies:
\begin{itemize}
\item $1 \le  l \le r \le m$
\item $T_i=f$,  \ \ \ \ \ \ \ $\forall i \in [l, r]$
\item $ T_l > T_{l-1}$, \ \ \ if $l>1$
\item $ T_r < T_{r+1}$, \ \ \ if $r<m$
\end{itemize}
There are at most $m$ blocks, which form a \emph{block set}. It fully captures the information in the sorted array $T$. An array of pointers called $PtrB$ is also needed to make a link from each element in $T$ to its relevant block. According to the definition of block, we always have: 
\begin{equation}
T_i = PtrB[i].f, \ \   \textrm{and} \ \   PtrB[i].l \le i \le PtrB[i].r ~ ~ .
\end{equation}
Here we use ``$.l$'' to denote the member $l$ of a block, and so on.
\begin{figure}[t]
\centering
\includegraphics[width=3.18in]{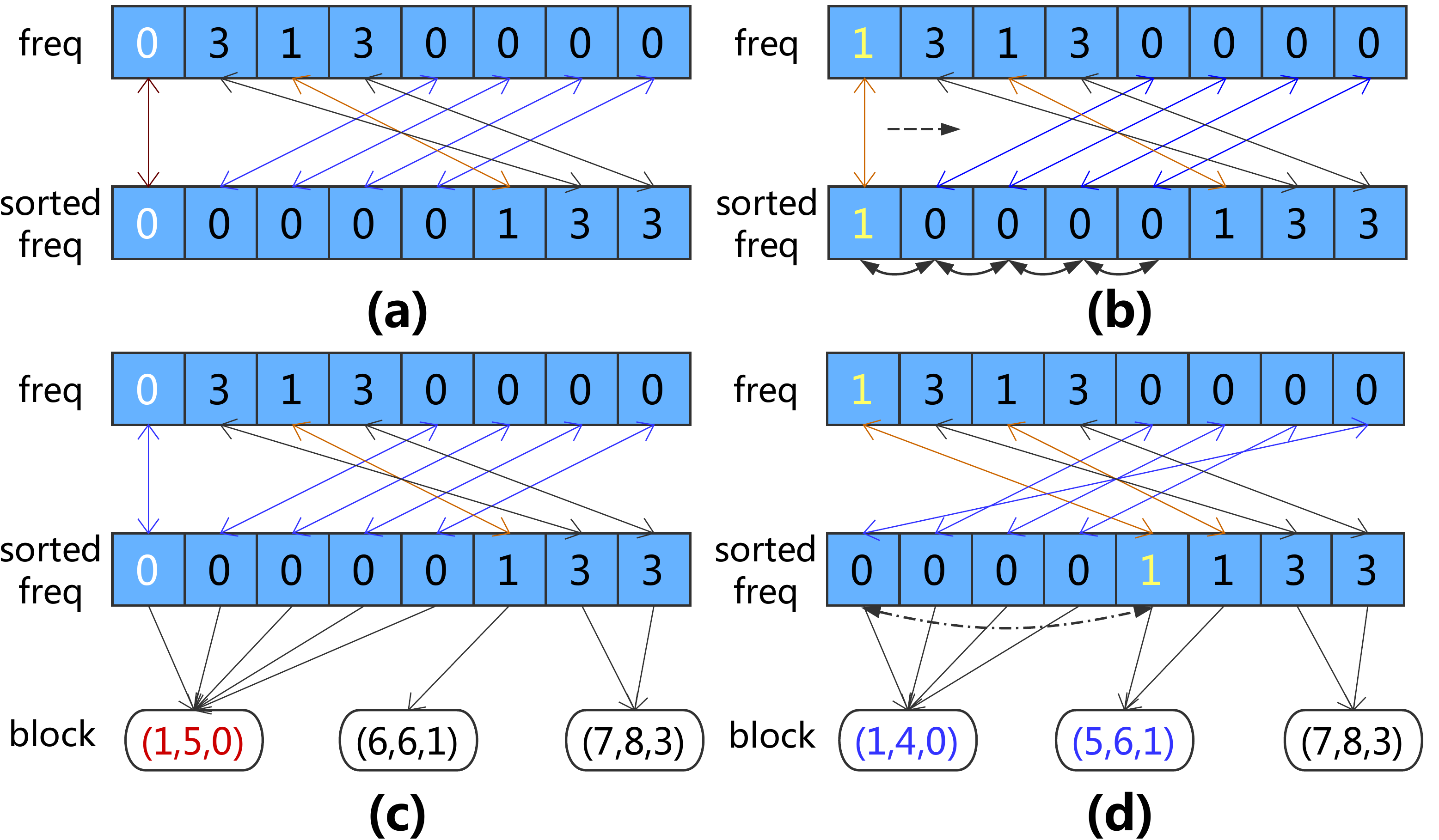}
\caption{Illustration of the proposed \emph{block set} for 
maintaining array $T$. (a) The initial $F$ and $T$. 
(b) When ``1''is added to $A$, a brute-force approach to maintain $T$ includes four swaps 
of ``1'' rightwards and updates of $FtoT$ and $TtoF$.
(c) The initial $F$ and $T$, and the \emph{block set}. $F$ and $T$ are up to be modified. 
(d) With the information of \emph{block}, the swapping destination in $T$ can be easily determined.} 
\label{fig:figure1}
\end{figure}

The block set $B$ represents the sorted frequency array $T$, while with arrays $FtoS$, $StoF$ and $PtrB$ we no longer need to store $F$ and $T$. These proposed new data structure well profile the dynamic array $A$. The remaining thing is to maintain them and answer the statistical queries on $A$ in an efficient manner.  


%

\subsection{S-Profile: the $O(1)$-Complexity Updating Algorithm}
We first consider the situation where an integer is added to $A$. As shown in Figure 1(a) and 1(b), a brute-fore approach to update $T$ is swapping the updated frequency to its right-hand neighbor one by one, until $T$ is in the appropriate order again. Now, with the proposed $PtrB$ and the block it points to, we can easily determine the index of $T$ which is the destination of swapping the updated frequency. Then, we can update the relevant two blocks and pointer arrays (see Figure 1(d)).


Now, based on the situation shown as Figure 1(d), we assume a ``4'' is removed from $A$. As shown in Figure 2, we first locate the updated element in $T$. 
\begin{figure}[b]
\centering
\includegraphics[width=3.18in]{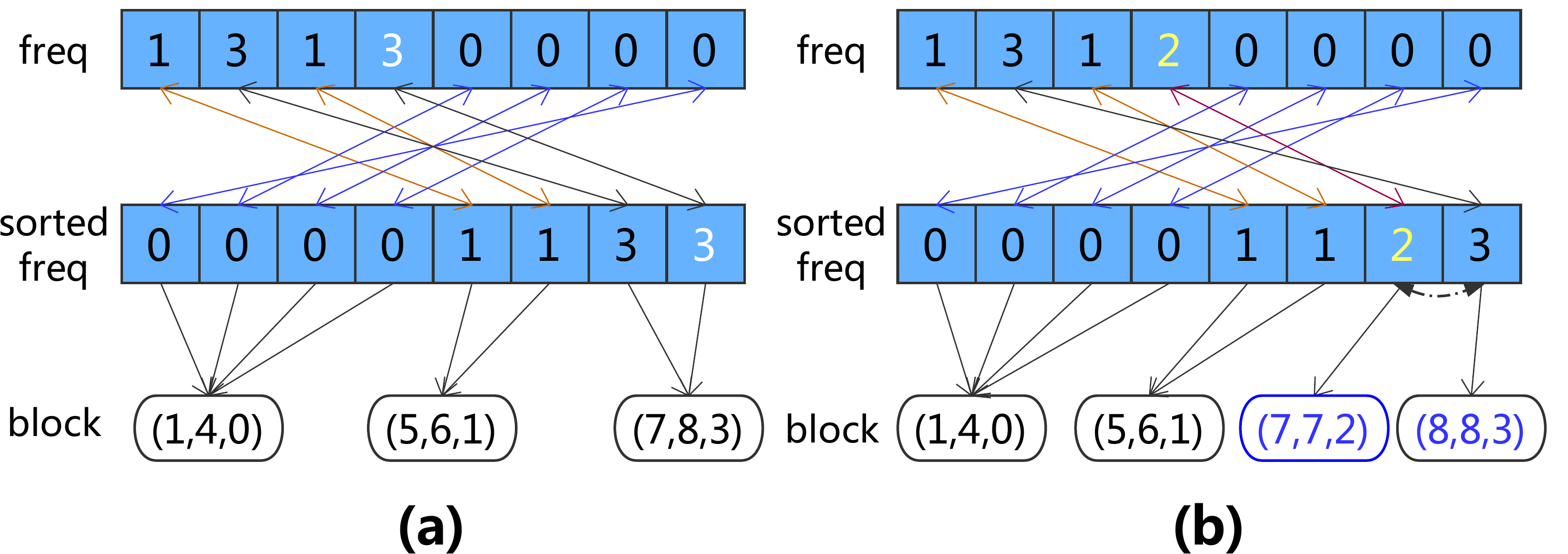}
\caption{Illustration of the proposed data structure for a ``remove'' action on $A$. (a) The initial $F$ and $T$, and the \emph{block set}. A ``4'' is going to be deleted from $A$.
(b) With the information of \emph{block} and the pointer arrays, the \emph{block set} can be easily updated to reflect the ordered $T$.}
\label{fig:figure2}
\end{figure}Then, with the information in its corresponding block we know with which it should be swapped. We further check if the updated frequency exists in $T$ before. If it does not we need to create a new block (the case in Figure 2(b)), otherwise another block is modified.

The whole details of the algorithm for updating the data structure and returning the mode of $A$ is described as Algorithm 1. We assume the data structures ($B$, $FtoS$, $StoF$ and $PtrB$) have been initialized, while Algorithm 1 responds to an event in the log stream and returns the updated mode and frequency.  
\begin{algorithm}
\caption{S-Profile for updating the mode of array}\label{alg:our}
\begin{flushleft}
\textbf{Input:}  A tuple ($x, c$) in log stream, block set $B$, pointer arrays $FtoT$, $TtoF$ and $PtrB$, the length of sorted frequency array $m$  \\
\textbf{Output:} The mode $M$, and the frequency $v$.
\end{flushleft}
\begin{algorithmic}[1]

\State $rank\gets FtoT[x]$
\State $b\gets PtrB[rank]$
\State $l\gets b.l$ ; \ \ \ \ $r\gets b.r$;
\If{$c$ is an ``add'' action}
\State $b.r \gets b.r-1$
\If{$b.r < b.l$}
\State Delete $b$
\EndIf
\If{$r<m$ \ and \ $b.f+1=PtrB[r+1].f$}
\State $PtrB[r]\gets PtrB[r+1]$
\State $PtrB[r].l\gets PtrB[r].l-1$
\Else
\State Create a new block in $B$ and assign it to $PtrB[r]$
\State $PtrB[r] \gets (r, \ r, \ b.f+1)$
\EndIf
\Else   \ \ \ \  \ \ \ \ \   /* It's a ``remove'' action */
\State $b.l \gets b.l+1$
\If{$b.r < b.l$}
\State Delete $b$
\EndIf
\If{$l>1$ \ and \ $b.f-1=PtrB[l-1].f$}
\State $PtrB[l]\gets PtrB[l-1]$
\State $PtrB[l].r\gets PtrB[l].r+1$
\Else
\State Create a new block in $B$ and assign it to $PtrB[l]$ 
\State $PtrB[l]\gets (l, \ l,\  b.f-1)$
\EndIf
\EndIf
\State $M \gets TtoF[PtrB[m].l\dots PtrB[m].r]$
\State $v \gets PtrB[m].f$
\end{algorithmic}
\end{algorithm}

As the proposed data structure maintains the sorted frequency array, it can be utilized to calculate the object with the minimum frequency (maybe a negative number) as well. We just need to replace Step 29 and 30 in Algorithm 1 with the following steps.

\begin{tabular}{l}
\toprule
29a: $M \gets TtoF[PtrB[1].l\dots PtrB[1].r]$\\ 
30a: $v \gets PtrB[1].f$ \\
\bottomrule
\end{tabular}

We can observe that the time complexity of the S-Profile algorithm is $O(1)$, as there is no iteration at all. The space complexity is $O(m)$, where $m$ is the maximum number of objects in the log stream. Precisely, it needs $3m$ integers to store the pointer arrays and an additional storage for $B$. In the worst case $B$ includes $m$ blocks, but usually this number is much smaller than $m$.

Other queries on statistics of objects can also be answered. For example, the top-K order element is that whose frequency is the K-th largest. We can just use  $PtrB[m-K+1]$ to locate  the block. Then, the frequency and object id can be obtained with block's member and the $TtoF$ array. Especially, the median in frequency can be located with the K/2-th element of the $PtrB$ array.

\subsection{Possible Applications}
For some mission-critical tasks (e.g., fraud detection) in big graphs, 
the efficiency to make decisions and infer interesting patterns is crucial. 
As a result, recent years have witnessed an increasing interest 
in heuristic ``shaving'' algorithms with low computational 
complexity~\cite{hooi2016fraudar, shin2017densealert}. 
A critical step of them is to keep
finding low-degree nodes at every time of shaving nodes from a graph.
Thus, S-Profile can be plugged into such algorithms for further speedup, by treating a node as an object and its degree as frequency.

Furthermore, S-Profile can also deal with a sliding window on a log stream, 
by letting every tuple $(x_i, c_i)$ outdated from the window be a new incoming
tuple $(x_i, \bar c_i)$, where $\bar c_i$ is the opposite action of $c_i$.



\section{Experimental Results}
We have implemented the proposed S-Profile algorithm and its counterparts in C++, and tested them with randomly generated log streams. The streams are produced with the following steps. We first randomly generate an "add" or "remove" action, with 70\% and 30\% probabilities respectively. Then, for each "add" action we randomly choose an object id according to a probability distribution (called \emph{posPDF}). For each "remove" action another distribution (called \emph{negPDF}) is used to randomly choose an object id. With this procedure, we obtained three test log streams:
\begin{itemize}
\item Stream1: both \emph{posPDF} and \emph{negPDF} are uniform random distribution on [1, m].
\item Stream2: both \emph{posPDF} and \emph{negPDF} are normal distributions with $\mu=2m/3, m/3$ and $\sigma=m/6, m/6 $, respectively.
\item Stream3: \emph{posPDF} is a normal distribution ($\mu=4m/5$, $\sigma= m$), while \emph{negPDF} is a lognormal distribution ($\mu=3m/5$, $\sigma= m$).
\end{itemize}

In the following subsections, we first compare the proposed S-Profile with the heap based approach, for updating the mode and frequency. Then, the comparison with the balanced tree is presented for calculating the median. All experiments are carried out on a Linux machine with Intel Xeon E5-2630 CPUs (2.30 GHz). The CPU time (in second) of different algorithms are reported.


\subsection{Comparison with the Heap}
Heap is a kind of binary tree where the value in parent node must be larger or equal to the values in its children. Used to maintain the sorted frequency array, it is easy to obtain the mode (the root has the largest frequency). Noticed the balanced tree  is inferior to the heap for calculating the mode. In Figure 3, 
\begin{figure}[b]
\centering
\includegraphics[width=3.18in]{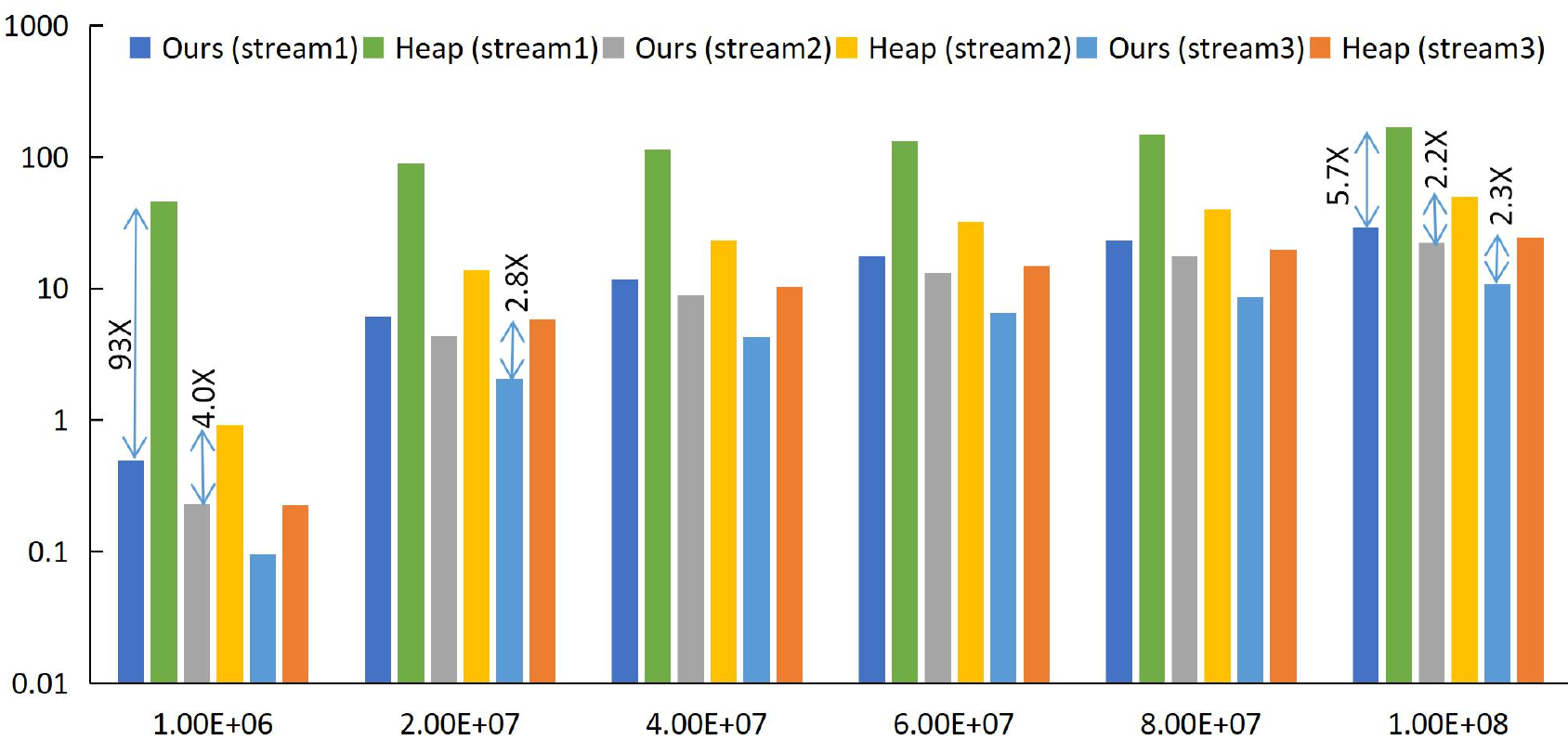}
\caption{CPU time of the heap based method and our method for calculating the mode ($m=10^8$).}
\label{fig:figure3}
\end{figure}we show the CPU time consumed for updating the mode with the heap based method and our S-Profile. The x-axis means the number of processed tuples ($n$). From the results we see that our method is at least 2.2X faster than the heap based method. Another experiment is carried out to fix $n=10^8$ while varying $m$. The results shown in Figure 4 also reveal that our S-Profile is at least 2X faster.

\begin{figure}[h]
\centering
\includegraphics[width=3.18in]{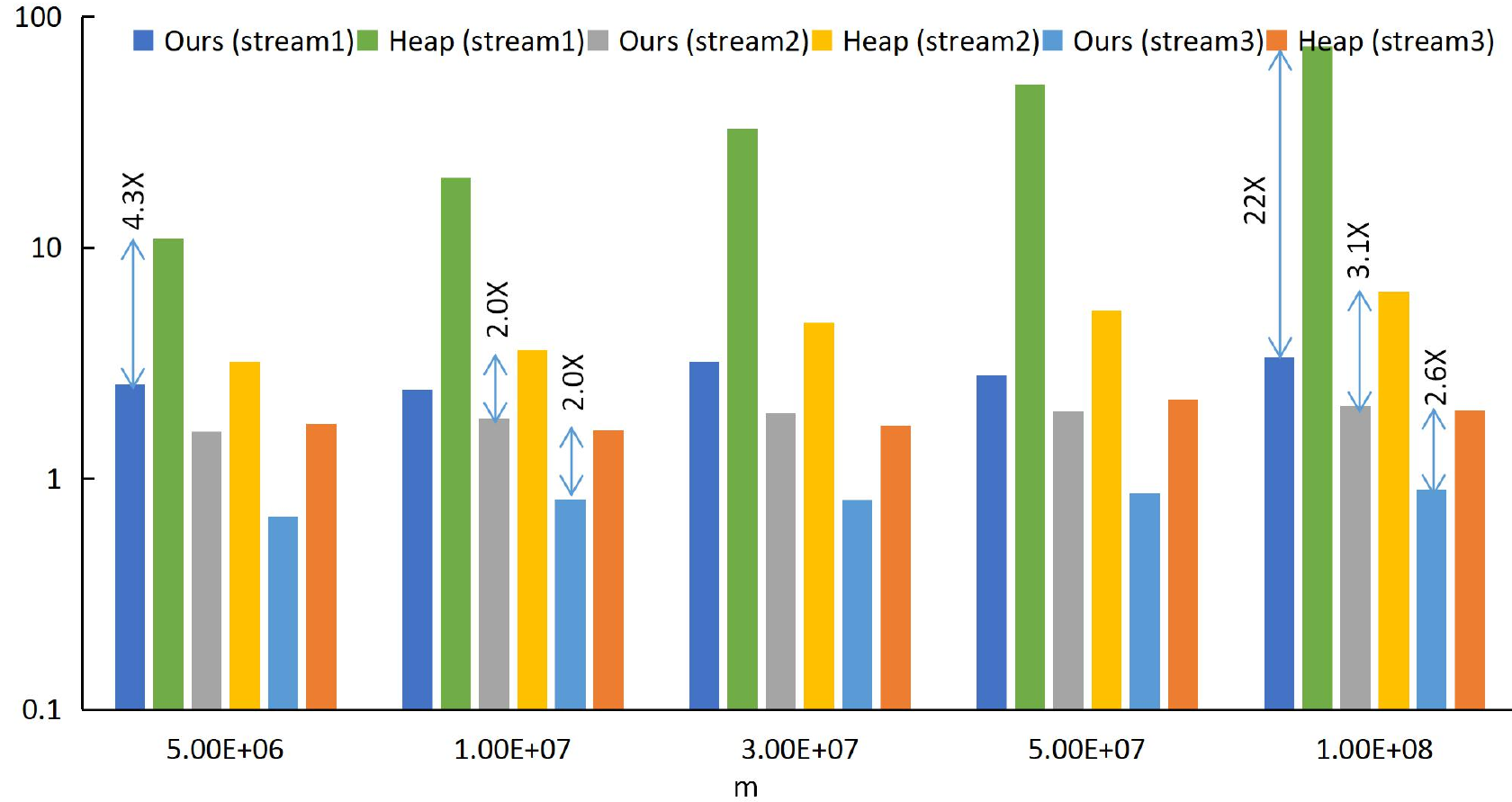}
\caption{CPU time of the heap based method and our method for calculating the mode ($n=10^8$).}
\label{fig:figure3b}
\end{figure}

\begin{figure}[h]
\centering
\includegraphics[width=2.1in]{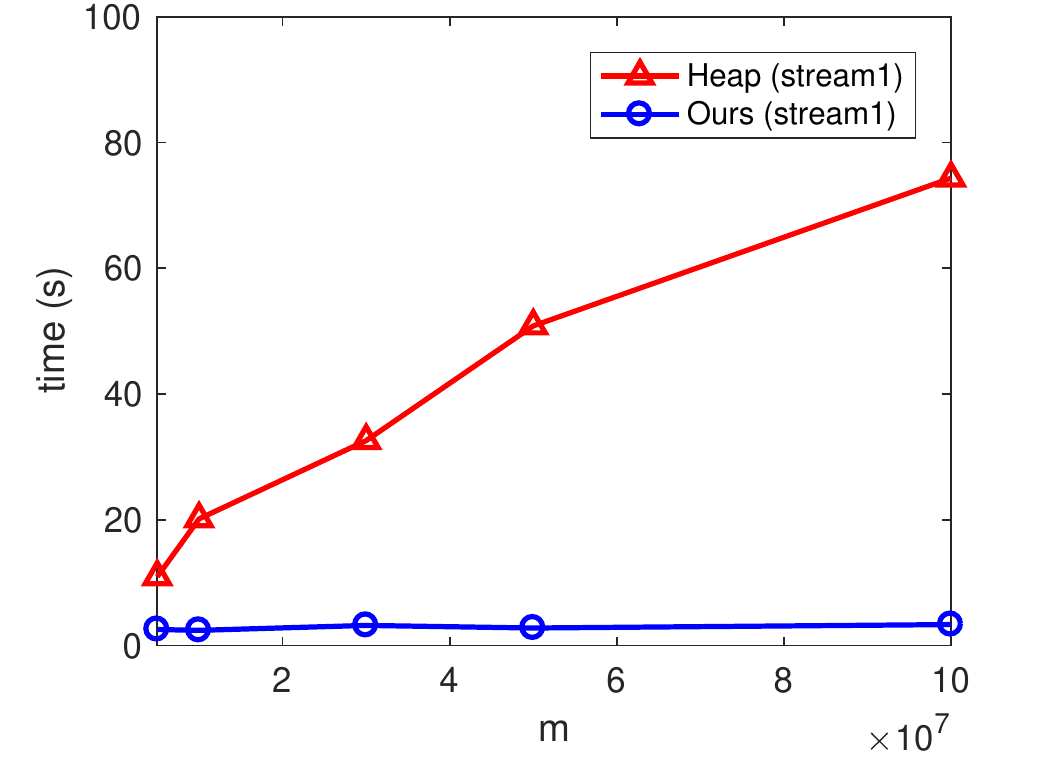}
\caption{The trends of CPU time for varied $m$ ($n=10^8$).}
\label{fig:figure3c}
\end{figure}
For different kind of log stream, the performance of the heap based method varies a lot. For the worst case updating the heap needs $O(\log m)$ time,  despite this rarely happens in our tested screams. On the contrary, S-Profile needs $O(1)$ time for updating the data structure. This advantage is verified by the rather flat trend shown in Figure 5. 

It should be emphasized that, in addition to  the speedup to the heap based method, our S-Profile possesses the advantage of wider applicability. Our method is not restricted to calculating the mode and corresponding frequency. As it well profiles the sorted frequency array, with it answering the queries on top-K and other statistics of objects is trivial and fast.

\subsection{Comparison with the Balanced Tree}

\begin{figure}[h]
\centering
\includegraphics[width=3in]{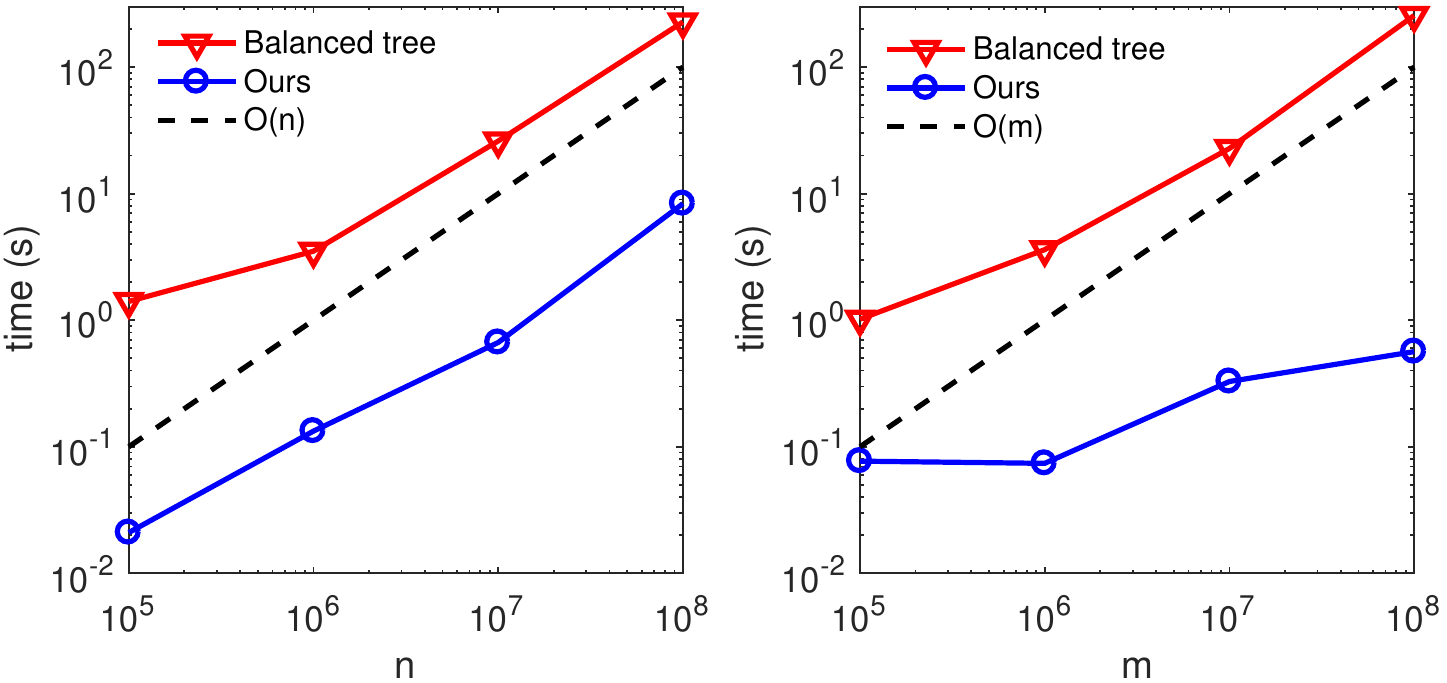}
\caption{Comparison of the balanced tree based method and the proposed algorithm for calculating the median. Left: Trend of CPU time vs. $n$ ($m=10^6$). Right: Trend of CPU time vs. $m$ ($n=10^6$).}
\label{fig:figure4}
\end{figure}

The proposed S-Profile can also calculate the median of the dynamic array. We compare it with the balanced tree based method implemented in the GNU C++ PBDS \cite{PBDS}, which is more efficient than our implementation of balanced tree.
The trends of CPU time are shown in Figure \ref{fig:figure4}. They show that the runtime of the proposed S-Profile increases much less than that of the balanced tree based method when $m$ increases. We can observe that the time of S-Profile is linearly depends on $n$, the number of modifications on array $A$, and hardly varies with different $m$. On the contrary, the balanced tree based method exhibits superlinearly increase whether with $n$ or $m$. Overall, the test results show that S-Profile is from 13X to 452X faster than the balanced tree based method on updating the median of the dynamic array.

\section{Conclusions}
We propose an accurate algorithm, S-Profile, to
fast keep profiling the dynamic array from online systems. It has the following advantages:
\begin{itemize}
    \item[-] \textbf{Optimal efficiency:} 
	S-Profile needs $O(1)$ time complexity for every updating of a dynamic array,  
    and totally linear complexity in memory. 
    \item[-] \textbf{Querying Statistics:}
	With profiling, we can answer the statistical queries in a trivial and fast way.
    \item[-] \textbf{Applicable:} S-Profile can be plugged into most of log
	streams, and heuristic graph mining algorithms. 
\end{itemize}


\end{document}